\documentstyle[12pt]{article}
\begin{document}
\begin{center}
{\bf Bacterial evolution and the Bak-Sneppen model}
\vskip 1cm
Indrani Bose and Indranath Chaudhuri$^\ddag$
\\Department of Physics, \\Bose Institute,
\\93/1, Acharya Prafulla Chandra Road,
\\Calcutta-700 009, India.
\\$\ddag$ Department of Physics, \\St. Xavier's College,
\\30, Park Street, Calcutta-700016, India.
\end{center}

\begin {abstract}
Recently, Lenski et al \cite{Elena,Lenski,Travisano} have carried out several experiments on bacterial evolution.
Their findings support the theory of punctuated equilibrium in biological
evolution. They have further quantified the relative contributions of adaptation,
chance and history to bacterial evolution. In this paper, we show that a modified
$M$-trait Bak-Sneppen model can explain many of the experimental results in a
qualitative manner.
\end{abstract}

Recently, a set of experiments has been carried out on bacterial evolution
which are part of a new sub-discipline in the area of evolutionary
biology, namely, experimental evolution \cite{Elena,Lenski,Travisano,RE}. This implies the study, in
the laboratory, of the fundamental processes of evolutionary change, namely,
spontaneous mutation and adaptation by natural selection. Experiments on
evolutionary dynamics require passage through thousands of generations which
is an impossibility for most living species. Microorganisms like bacteria, yeast
or viruses, however, have very short generation times (bacteria like E.coli have
about seven generations everyday in sugar solution). The short generation times
make it possible to observe population dynamics over thousands of generations 
and thereby address a wide range of evolutionary questions. Darwin laid the
foundation of evolutionary biology by setting forth the principle of adaptation
by species through natural selection \cite{Darwin}. It is now known 
that spontaneous mutation plays an important role in generating differences
among individual organisms. A mutation involves a change in the base sequence 
of a DNA and can occur with a certain probability during cell division.  Mutations are random
events and may be harmful, neutral or beneficial as regards their effect on
an organism. According to the modern version of Darwin's theory, random mutations
give rise to heritable differences among organisms whereas natural selection
tends to increase the number of fitter variants. The processes of reproduction,
mutation and natural selection are responsible for evolution of species or that
of a single population.

The experiments on the evolutionary dynamics of the bacterial population
E.coli have been  carried out by Lenski and co-workers
\cite{Elena,Lenski,Travisano}. As pointed out by them
\cite{T}, a large population size of E.coli ensures that a large number of mutations
occur in every generation so that the origin and the fate of genetic variation can 
be well studied. It is possible to store bacteria in suspended animation at low
temperatures. One can then measure the relative fitness (RF) of descendant and
ancestral populations by placing them in direct competition. The RF is expressed 
as the ratio of the realized growth rates of the two populations. One can also measure
a morphological quantity like the average cell size which is preserved in fossil
records. Finally, the populations are easy to handle and propagate so that
intensive replication of experiments is possible allowing subtle effects
to be measured. Bak and Sneppen (BS)\cite{Bak,B} have proposed a model of
evolution at the level of the global ecosystem of interacting species (macroevolution).
The model self-organises into a critical steady state characterised by
power-laws of various types. The major focus of the BS model is on
self-organised-criticality (SOC) and its various aspects. In this paper,
we apply the BS model to evolution at the level of a single population of
reproducing bacteria (microevolution) and show that the evolutionary dynamics of
the BS model can describe many of the experimental results of Lenski et al
in a qualitative manner.

In the first experiment \cite{Elena,Lenski}, Lenski and co-workers studied an evolving bacterial population for
approximately 10,000 generations. The bacterial population was allowed to expand
to $5\times10^{8}$ cells in low sugar solution. At the end of the growth period (one day),
$1/100$ th of the population was siphoned into a fresh flask of food to allow
the population to evolve. Since there was a 100-fold expansion of the bacterial
population in a day, $6.6 (=\log_{2}{100})$ generations of
binary fission occurred during this time. Every fifteen days, a sample of the 
population was frozen for later analysis giving rise to a `frozen fossil record'.
Since the population originated from a single cell, mutations, about $10^{6}$ everyday,
provided the sole source of genetic variation. Four years later, Lenski et al
had data for the evolving bacterial population over 10,000 generations.
They measured two quantities, the average cell size 
and the RF. They found that the average cell size
and the RF grow in a punctuated manner, i.e., in steps, as a function
of time (number of generations) when the data are plotted every 100 generations
(inset of Fig.1). At a larger interval of 500 generations, the changes 
appear to be gradual (inset of Fig.2 ). A major debate in evolutionary
biology revolves around the question of whether evolution is best described as
a gradual change or occurs in bursts. In the latter case, short periods of 
evolutionary activity are punctuated by long periods of stasis. This is the
theory of punctuated equilibrium (PE). Lenski et al's data seem to support
this theory though their interpretation is open to controversy \cite{Coyne}. The
experiment, however, clearly demonstrates that  both the average cell
size and the RF of the evolving population grow over a certain time interval. In the first 2000
generations or so, there is a rapid growth followed by a period of slower
growth till the growth is imperceptible. The bacteria, being in low sugar 
solution, have to compete for the food. Natural selection favours the 
mutations that confer some competitive advantage in exploiting the experimental
environment. This leads to adaptation of the bacterial population to the 
environment through the emergence of larger and fitter variants.

In the second experiment \cite{Lenski}, twelve populations of E.coli were evolved over 10,000
generations in identical environments. Each population was founded by a single
cell from an asexual clone to eliminate genetic variation within and between
populations. It was found that the replicate populations, after 10,000 generations,
differ considerably from one another in both the average cell size and the RF 
(inset of Fig.3 ), even though the populations evolved in identical
environments. In the third experiment \cite{Travisano}, the relative contributions of adaptation,
chance (mutations) and history to evolution were investigated. Twelve replicate
populations were founded from a single clone of E.coli and serially propagated
for 2000 generations in glucose-limited medium. The 12 populations had similar
fitness values when evolving in glucose medium but when put into a maltose-limited
medium showed large differences in fitness values. Some populations thrived 
while some others were found to languish. Ancestral fitness values of the 
populations in maltose were thus very heterogeneous (inset of Fig.4). One 
genotype from each of the 12 replicate populations was cloned and from this 
3 new replicate populations were founded. The 36 populations were then evolved 
under ancestral conditions in the maltose medium. The inset in Fig.4 shows
the derived fitness in maltose versus the ancestrial fitness in maltose for 36
populations.

Boettcher and Paczuski \cite{Bo} have generalised the BS model to the $M$-trait
BS model in which each species is characterised by $M$ traits rather than one
as in the original BS model. The rules of evolutionary dynamics are, however,
the same. We now show that the $M$-trait ($M =2$) BS model with a minor variation
in the rules, can reproduce some of the experimental observations of Lenski
et al. The model is applied to a single, evolving population of E.coli, rather
than to many species. The BS model gives a coarse-grained representation of real
evolution but contains the essential elements to capture the course of evolution. 
In our modified $M=2$ BS model, the bacterial population is divided into $N$ categories.
Each category contains bacteria of similar characteristics. The $N$ categories 
correspond to the $N$ sites of a one-dimensional (1D) lattice with periodic 
boundary conditions. In the original BS model, each site represents a species.
Two traits, namely, cell size and fitness are associated with the population at
each site $i, i=1,2,...,N$. One assigns a number (between 0 and 1) to each of the
traits at all the $N$ sites. At each time step, the two sites with the minimum 
values for each of the two traits are identified. Mutations occur to bring about
changes in the traits. The minimum random numbers are replaced by new random
numbers. This takes into account the fact that the weakest species are the
most liable to mutate.
In the original $M=2$ BS model, the minimum value, amongst all the $2N$
values of the two traits, is replaced by a new random number. The random numbers
associated with any one of the traits of the neighbouring sites are also replaced
by new random numbers. This is to take into account the linkage of neighbouring
populations in food chain. The bacteria, evolving in low sugar solution, have
to compete for food and one assumes that the neighbouring sub-populations
affect each other the most in the evolution of traits.
The last two steps are repeated and averages are taken
for both the traits locally (over 40 sites) and globally (over 2000 sites).
The minor change in the evolutionary rules from those of the $M=2$ BS model gives a better agreement with the
experimental results. In fact, other minor variations of the rules (like
changing the random number interval from (0-1) to a smaller range of values)
may provide an improved agreement. The essential ingredients of the BS model
are, however, retained.
Unlike in the original BS model, we calculate quantities from the very beginning
and not after the SOC state is reached. The SOC state corresponds to the region
in which evolutionary growth is imperceptible and fluctuates around an 
average value. Our major focus is on the evolutionary dynamics leading to the
critical state as this dynamics has been probed experimentally.
Fig.1  shows the variation of the
RF  versus time. The inset shows the experimental data 
\cite{Elena,Lenski}.
The averages are taken over 40 sites and every 100 time steps. The local averaging
gives rise to an improved quality of data points. Fig. 2
shows the corresponding variation with averages taken every 500 time steps
and over the whole lattice.  For a very large lattice one needs to take only global averages.
The RF is defined to be the ratio of the current 
fitness and the initial fitness at time $t=0$. In the actual experiment, fitness is
related to the growth rate of the bacterial population via replication.  
The RF  increases rapidly during the first 2000 
generations. After that the growth becomes slower till it becomes imperceptible. 
The rapid growths can be fit by an hyperbolic model
\begin{eqnarray}
y=x_{0}+\frac{ax}{b+x}
\end{eqnarray}
for both the experimental and simulation data. During the periods of punctuation
the beneficial mutations have no significant effect. When such mutations
occur in quick succession, rapid evolutionary growth is observed. Recent research findings \cite{VM} have highlighted
the importance of large beneficial mutations in the initial stages of evolutionary
growth. Organisms must adapt to the new conditions fairly quickly in order to
survive. Later, mutations with smaller effect fine-tune the adaptation. Fig.2
shows this clearly with a rapid evolutionary growth in the first 2000 generations
brought about by beneficial mutations of large effect. The growths are 
imperceptible when the bacterial population gets adapted to its environment.
The average cell size as a function of time has similar variations as in the case
of the RF (Figs. 1 and 2).

Fig.3 shows a comparison of the simulation data for the RF with the data (inset) 
of the second experiment of Lenski et al \cite{Lenski}. The plots show that the 
independent populations diverge significantly from one another. In the simulation, the initial random number
seed was chosen to be different for the six populations. The average fitness
 at time $t=0$ does not vary noticably from one
population to another. Fig.4 shows the simulation results for the third 
experiment \cite{Travisano} with $2\times4$ populations rather than the $3\times12$ populations
in the actual experiment. In the experiment, the populations growing in glucose-
limited medium were transferred after 2000 generations to maltose-limited medium.
In the latter medium, the average fitness values of the populations showed
large differences. Thus, in the simulation for the maltose medium, one starts
with widely different average fitness values for the populations. The populations
are evolved for 1000 time steps. One finds, in agreement with the experimental
results, that after 1000 generations, the average fitness values have similar
magnitudes. This shows that adaptation and chance (effect of mutation) have
eliminated the initial heterogeneity in average fitness values to a great
extent. The effect of history (initial heterogeneity) is reduced after several 
generations of evolution. The effect of adaptation is pronounced (shown by the
evolution of the data points above the isocline). The effect of chance is seen
in the small dispersion in the average fitness values of the two populations
corresponding to each genotype.

Lenski et al \cite{Rose,Gerrish,Johnson,Vasi} have developed theories based on
standard population-genetics approaches to explain some of their experimental results.
Such theories provide a microscopic picture of evolution but require detailed
information about various parameters. These include the Malthusian parameter
$m_{i}$of a strain ($m_i= ln[N_i(1)/N_i(0)]$ per day, where $N_i(0)$ and
$N_i(1)$ are the densities of the population at the beginning and the end of
the day), the fitness $w_{ij}$ of one strain relative to another ($w_{ij}= m_i/m_j$),
the selection coefficient $S_{ij}$ ($S_{ij} = w_{ij}- 1$) and the selection
rate constant $r_{ij}$ ( $ = m_i - m_j = m_j S_{ij}$). When a mutation occurs
for the first time, the frequency of the mutant genotype, $P(0)$, is equal to
$1/N$, where $N$ is the population size. If the new mutation is not lost by
drift, the rate of change in the frequency of the allele (genetic variant)
is governed by the equation \cite{Rose}
\begin{eqnarray}
\frac{dP}{dt} = r_{ij}P(1-P)
\end{eqnarray}
where $r_{ij}$, the selection-rate constant, is the difference
in the Malthusian parameters of the favourable mutant and its progenitor.
Mean fitness in the population, $W_{av}(t)$, depends on the frequency of the favoured mutant according to
\begin{eqnarray}
W_{av}= 1 + S_{ij} P(t) \cong 1 + r_{ij} P(t)/m_{av}
\end{eqnarray}
where $m_{av}$ is the average Malthusian parameter. Solving Eqs. (2) and (3)
for appropriate values of the parameters, Lenski et al could reproduce the
step-like trajectory for relative fitness versus time, observed in
experiments (inset of Fig. 1). The beneficial mutation takes many generations
to reach a frequency which has appreciable effect on relative fitness and
hence the plateau in the trajectory.
Population genetics approaches certainly give a more detailed and accurate
picture of biological evolution. Such approaches, however, require detailed assumptions
and considerable computational effort. The $M=2$ BS model, on the other hand, is
an oversimplified model which incorporates the essential features of real evolution
in the form of a set of rules. In this paper, we have shown that the minimal model
gives a satisfactory description of the experimental data on bacterial
evolution. Both simulation and experiments show evidence of PE on a short
time scale (data points taken every 100 generations). Over a longer time scale,
both show a hyperbolic growth in relative fitness. This is also true for the
average cell size. Recent exhaustive studies of fossil beds lend support to
the theory of PE \cite{Kerr}. The simulation further shows that chance
(mutations) gives rise to parallel evolution of replicate populations.
The relative contributions of adaptation, chance and history to average fitness
before and after 1000 generations in maltose are correctly highlighted in
the simulation data. Thus a BS-type model can explain the major experimental
observations on bacterial evolution in a qualitative manner.

The BS model is well-known for its characterization of the self-organised
critical state in the ecosystem of interacting species. Such self-organization
can also occur in an evolving population. In the case of E.coli, the critical
state is obtained when evolutionary growth becomes imperceptible and
fluctuates around a steady value. With appropriate designing of experiments,
the phenomenon of SOC in an evolving bacterial population can be studied experimentally.
Experiments have recently been performed on the
growth of RNA virus fitness \cite{No}. The adaptive evolutionary capacity in
this case is overwhelming. The gain in fitness is nearly 5000
after 50 passages. In that time, the gain in E.coli fitness changes can be 
explained by an hyperbolic model whereas RNA virus evolution 
follows exponential kinetics. Again, a simple model has been proposed to explain
the experimental observations. The several experiments on the evolutionary
dynamics of microorganisms open up the possibility of a rich interplay
between theory and experiments. Simple models like the BS model capture
the significant features of evolutionary dynamics of single populations. It
is, however, desirable to explore the connection between such models and the
more comprehensive population-genetics approaches.

\section*{Acknowledgement}
We are greatful to R.E.Lenski for sending us his publications on bacterial
evolution. We also acknowledge the help from the Distributed Informatics
Centre, Bose Institute, in preparing the manuscript.

\newpage
\section*{Figure Captions}
\begin{description}
\item[Fig.1] Relative fitness versus time in experiment \cite{Lenski} and simulation.
A local average is taken over 40 sites in simulation. The experimental data
points are taken every 100 generations.
\item[Fig.2] Relative fitness versus time in simulation and experiment (inset)
\cite{Lenski}. The data points are taken every 500 generations and the average
is over the 2000 sites of the lattice.
\item[Fig.3] Trajectories for relative fitness in six replicate populations
of bacteria during 10,000 generations (simulation) and the same for twelve
replicate populations (experiment). The data points are taken at an interval
of 500 generations.
\item[Fig.4] Evolution of fitness during 1000 generations in maltose. Derived
versus ancestral values for relative fitness in 8 populations (simulation)
and the same for 36 populations (experiment). The different symbols indicate
the different progenitor genotypes.
\end{description}

\end{document}